\documentstyle[prl,aps,epsf,floats]{revtex}
\def\narrowtext{} \tighten \twocolumn
\input epsf.sty

\begin{document}
\draft

\title{Renormalization of spectral lineshape and dispersion
below $T_c$ in Bi$_2$Sr$_2$CaCu$_2$O$_{8+\delta}$}
\author{
        A. Kaminski,$^{1,2}$
        M. Randeria,$^{3}$
        J. C. Campuzano,$^{1,2}$
        M. R. Norman,$^{2}$
        H. Fretwell,$^{4}$
        J. Mesot,$^{5}$
        T. Sato,$^6$
        T. Takahashi,$^6$
        K. Kadowaki$^7$
       }
\address{
         (1) Department of Physics, University of Illinois at Chicago,
             Chicago, IL 60607\\
         (2) Materials Sciences Division, Argonne National Laboratory,
             Argonne, IL 60439 \\
         (3) Tata Institute of Fundamental Research, Mumbai 400005,
             India\\
         (4) Department of Physics,University of Wales Swansea,
             Swansea SA2 8PP, UK\\
         (5) Laboratory for Neutron Scattering, ETH Zurich and PSI
             Villigen,
             CH-5232 Villigen PSI, Switzerland\\
         (6) Department of Physics, Tohoku University,
             980 Sendai, Japan\\
         (7) Institute of Materials Science, University of Tsukuba,
             Ibaraki 305, Japan\\
         }
\address{%
\begin{minipage}[t]{6.0in}
\begin{abstract}
Angle-resolved photoemission (ARPES) data in
the superconducting state of Bi$_2$Sr$_2$CaCu$_2$O$_{8+\delta}$
show a kink in the dispersion along the zone diagonal, which
is related via a Kramers-Kr\"onig analysis to a drop in the
low-energy scattering rate. As one moves towards $(\pi,0)$,
this kink evolves into a spectral dip. The occurrence of these
anomalies in the dispersion and lineshape throughout the zone
indicate the presence of a new energy scale in the superconducting
state.
\typeout{polish abstract}
\end{abstract}
\pacs{74.25.Jb, 74.72.Hs, 79.60.Bm}
\end{minipage}}

\maketitle
\narrowtext

The high temperature superconductors exhibit many unusual properties,
one of the most striking being the linear temperature dependence of the
normal state resistivity. This behavior
has been attributed to the presence of a quantum critical point, where
the only relevant energy scale is the temperature\cite{VARMA}. However,
new energy scales become manifest below $T_c$ due to the appearance of
the superconducting gap and resulting collective excitations.
The effect of these new scales on the ARPES spectral function
below $T_c$ have been well studied near the $(\pi,0)$ point of the
zone \cite{MODE1,MODE2}.
In this Letter we show how these scales manifest themselves in
the spectral functions over the entire Brillouin zone.

Remarkably, we find that these effects are manifest
even on the zone diagonal where the gap vanishes,
with significant changes in both the
spectral lineshape and dispersion below $T_c$,
relative to the normal state (where the nodal points exhibit
quantum critical scaling \cite{VALLA}).
Specifically, below $T_c$ a kink in the dispersion develops along 
the diagonal at a finite energy ($\sim$70 meV). This 
is accompanied, as required by Kramers-Kr\"onig relations, by a 
reduction in the linewidth leading
to well-defined quasiparticles.  As one moves
away from the node, the renormalization
increases, and the kink in dispersion along the diagonal smoothly
evolves into the spectral dip \cite{MODE1}, with the same characteristic
energy scale throughout the zone.
We suggest that a natural interpretation of all of these
spectral renormalizations is in terms of the electron interacting
with a collective excitation below $T_c$, which is
likely that seen directly by neutron scattering
\cite{KEIMER}.

We begin our analysis by recalling \cite{NK}
that, within the impulse approximation, the
ARPES intensity for a quasi-two-dimensional system
is given by \cite{RESOLUTION}
$I({\bf k},\omega)= I_0({\bf k}) f(\omega)A({\bf k},\omega)$.
Here ${\bf k}$ is the in-plane momentum,
$\omega$ is the energy of the initial state relative to
the chemical potential, $f$ is
the Fermi function, $I_0$ is proportional
to the dipole matrix element $\left| M_{fi} \right|^2$,
and $A$ is the one-particle spectral function.
Fig.~1 shows data \cite{SAMPLE} as a function of ${\bf k}$ and $\omega$.

\begin{figure}[!t]
\epsfxsize=3.2in
\epsfbox{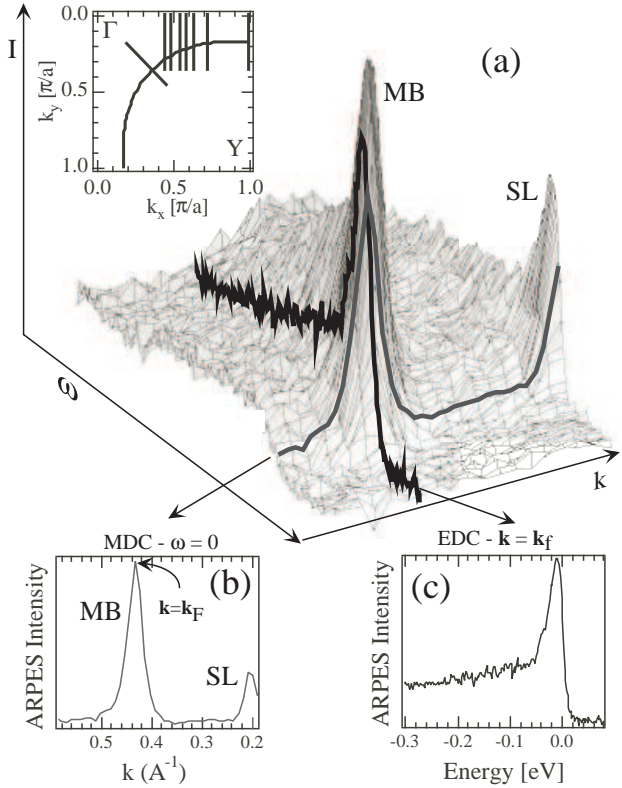}
\vspace{0.2cm}
\caption{
(a) The ARPES intensity as a function of ${\bf k}$ and $\omega$ at
$h\nu$=22eV and T=40K. MB is the main band, and SL a superlattice image.
(b) A constant $\omega$ cut (MDC) from (a).
(c) A constant ${\bf k}$ cut (EDC) from (a).
The diagonal line in the zone inset shows the location of the ${\bf k}$
cut; the curved line is the Fermi surface.
}
\label{fig1}
\end{figure}

$A({\bf k},\omega) = (-1 / \pi)\Im mG({\bf k},\omega + i0^{+})$
can be written as
\begin{equation}
A({\bf k},\omega) = {1 \over \pi}
{{\left| {\Sigma^{\prime\prime}({\bf k},\omega)} \right|} \over
{\left[{\omega-\epsilon_{{\bf k}}-\Sigma^{\prime}({\bf
k},\omega)}\right]^2
+ \left[{\Sigma^{\prime\prime}({\bf k},\omega)} \right]^2}}.
\label{spectral}
\end{equation}
where
the self-energy $\Sigma = \Sigma^{\prime}+i\Sigma^{\prime\prime}$
and $\epsilon_{{\bf k}}$ is the bare dispersion.
For ${\bf k}$ near $k_F$,
and varying normal to the Fermi surface (shown in
the inset in Fig.~1), we may write $\epsilon_{{\bf k}} \simeq v_F^0(k-
k_F)$,
where both $k_F(\theta)$ and the bare Fermi velocity $v_F^0(\theta)$
depend in general on the angle $\theta$ along the Fermi surface.

In Fig.~2a, we plot the dispersion of the spectral peak above $T_c$
obtained from constant ${\bf k}$ scans (energy distribution curves or
EDCs),
and the peak in momentum obtained from constant $\omega$ scans
(momentum distribution curves or MDCs) \cite{VALLA}
from data similar to Fig.~1.
We find that the EDC and MDC peak dispersions
are very different, a consequence of the $\omega$ dependence of
$\Sigma$.
To see this, we note from Eq.~(\ref{spectral}) that the
MDC at fixed $\omega$ is a Lorentzian centered at
$k = k_F + \left[\omega-\Sigma^{\prime}(\omega)\right]/v_F^0$,
with a width (HWHM) $W_M = |\Sigma^{\prime\prime}(\omega)|/v_F^0$,
{\it provided} (i) $\Sigma$ is essentially independent \cite{SIGMA}
of $k$ normal to the Fermi surface, {\it and} (ii) the dipole matrix 
elements do not vary significantly with $k$
over the range of interest.  That these two conditions are
fulfilled can be seen by the nearly Lorentzian
MDC lineshape of the data in Fig.~1b.

On the other hand, in general, the EDC at fixed ${\bf k}$ (Fig.~1c) 
has a non-Lorentzian lineshape reflecting the non-trivial
$\omega$-dependence of $\Sigma$, in addition to the Fermi cutoff at 
low energies. Thus the EDC peak is {\it not} given by
$\left[\omega- v_F^0(k-k_F) - \Sigma^{\prime}(\omega)\right] = 0$
but also involves $\Sigma^{\prime\prime}$, unlike the MDC peak.
Further, if the EDC peak is sharp enough, making a Taylor expansion 
we find that its width (HWHM) is given by $W_E \simeq |\Sigma^{\prime\prime}
(E_k)|/[1 - \partial\Sigma^{\prime}/\partial\omega|_{E_k}]$, 
where $E_k$ is the peak position.

We see that it is much simpler to interpret
the MDC peak positions, and thus focus on the
change in the MDC dispersion going from the normal (N) to the
superconducting (SC) state shown in Fig.~2b.
The striking feature of Fig.~2b is the development of a kink
in the dispersion below $T_c$.
At fixed $\omega$ let the dispersion change from $k_N$ to $k_{SC}$.
Using $v_F^0(k_N - k_{SC}) = \Sigma_{SC}^{\prime}(\omega)
- \Sigma_N^{\prime}(\omega)$, we directly obtain the change in
real part of $\Sigma$ plotted in Fig.~2c. The
Kramers-Kr\"onig transformation of $\Sigma_{SC}^{\prime} -
\Sigma_N^{\prime}$
then yields $\Sigma_{N}^{\prime\prime} - \Sigma_{SC}^{\prime\prime}$,
plotted in Fig.~2d, which shows that
$|\Sigma_{SC}^{\prime\prime}|$ is smaller than
$|\Sigma_N^{\prime\prime}|$
at low energies.

\begin{figure}[!t]
\epsfxsize=3.2in
\epsfbox{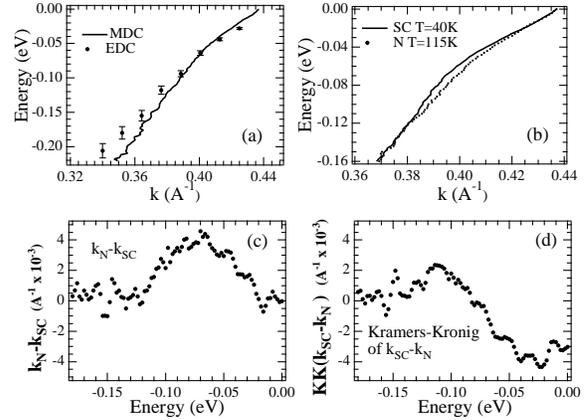}
\vspace{0.2cm}
\caption{
ARPES data along the $(\pi,\pi)$ direction at $h\nu$=28eV.
(a) EDC dispersion in the normal state compared to the MDC dispersion.
The EDCs are shown in Fig.~3d.
(b) MDC dispersions in the superconducting state (T=40K) and
normal state (T=115K). (c) change in MDC dispersion from (b).
(d) Kramers-Kr\"onig transform of (c).
}
\label{fig2}
\end{figure}

We compare these results in Fig.~3a with the
$W_M = |\Sigma^{\prime\prime}|/v_F^0$ estimated directly from the
MDC Lorentzian linewidths.
The normal state curve was obtained
from a linear fit to the corresponding MDC width data points in Fig.~3a,
and then the data from Fig.~2d was added to it to generate the
low temperature curve.
We are thus able to make a quantitative connection between the
appearance of a kink in the (MDC) dispersion below $T_c$ and
a drop in the low energy scattering
rate in the SC state relative to the normal state,
which leads to the appearance of quasiparticles below $T_c$
\cite{ADAM}.
We emphasize that we have estimated these $T$-dependent changes in the
complex self-energy without making fits to the EDC
lineshape, thus avoiding the problem of modeling the $\omega$ dependence
of $\Sigma$ and the extrinsic background.

In Fig. 3b, we plot the EDC width obtained as explained in \cite{ADAM} 
from Fig.~3d.  As an interesting exercise, we present in Fig.~3c the 
ratio of this EDC width to the MDC width of Fig.~3a (dotted lines), 
and compare it to the renormalized MDC velocities, 
$1/v\equiv dk/d\omega$, obtained directly by numerical differentiation 
of Fig.~2b (solid lines). We note that only for a sufficiently narrow 
EDC lineshape is the ratio $W_E/W_M \simeq v_F^0 / 
[1 - \partial\Sigma^{\prime}/\partial\omega] = v_F$. Interestingly, 
only in the SC state below the kink energy do these two quantities agree, 
which implies that only in this case does one have a Fermi liquid.

Similar kinks in the dispersion have been seen by ARPES in normal metals 
due to the electron-phonon interaction\cite{PHONON}. Phonons cannot be 
the cause here, since our kink disappears above $T_c$. Rather, our effect 
is suggestive of coupling to an electronic collective
excitation which only appears below $T_c$.

\begin{figure}[!t]
\epsfxsize=3.2in
\epsfbox{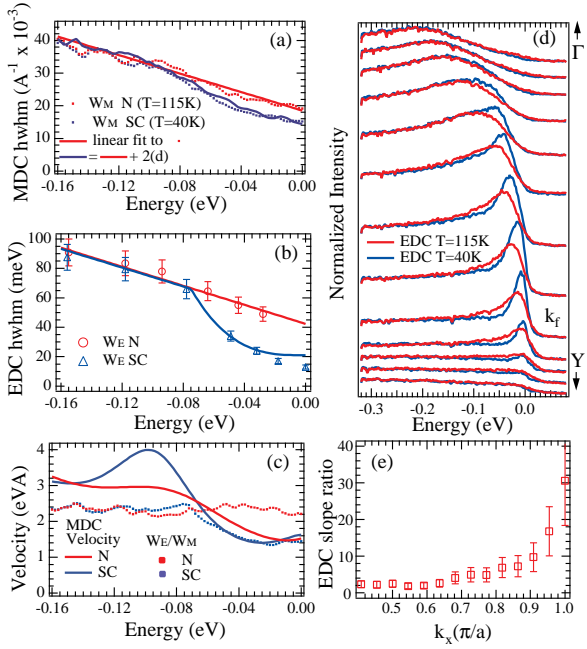}
\vspace{0.2cm}
\caption{
(a) Comparison of change in $\Sigma^{\prime\prime}$ obtained directly
from the MDC widths (HWHM) to the one obtained from the dispersion in 
Fig.~2d by using the Kramers-Kr\"onig transform.  (b) HWHM width obtained 
from EDCs shown in (d). Lines marked by fit are linear in normal state 
and linear/cubic in superconducting state.  The data in (b) fall below 
the fits at low energies because of the Fermi cut-off of the EDCs.
(c) Renormalized MDC velocity obtained from differentiating Fig. 2b 
(solid lines), compared to the ratio $W_{E}/W_{M}$ from (a) and (b). 
(e) Ratio of EDC dispersion slopes above and below the kink energy at various
points along the Fermi surface (from middle panels of Fig.~4).}
\label{fig3}
\end{figure}

We now study how the lineshape and dispersion evolve as we
move along the Fermi surface. Away from the node a quantitative
analysis (like the one above) becomes more complicated
\cite{COMPLICATIONS}
and will be presented in a later publication.
Here, we will simply present the data. In Fig.~4,
we plot raw (2D) data as obtained from our detector for a series of
cuts parallel to the $MY$ direction (normal state in left panels,
superconducting state in middle panels). We start from
the bottom row that corresponds to a cut close to the node and reveals
the same kink described above. As we move
towards $(\pi,0)$, the dispersion kink (middle panels) becomes more
pronounced and
at around $k_{x}$=0.55 develops into a break separating the
faster dispersing high energy part of the spectrum from the slower
dispersing low energy part. This break leads to the
appearance of two features in the EDCs, shown in the right panels of
Fig.~4. Further towards $(\pi,0)$, the low energy feature,
the quasiparticle peak, becomes almost dispersionless.
At the $(\pi,0)$ point, this break effect becomes the most pronounced,
giving rise to the well known peak/dip/hump \cite{MODE1} in the EDC.
We note that there is a continuous evolution in the zone from kink to
break, and these features all occur at exactly the same energy.

\begin{figure}[!t]
\epsfxsize=3.2in
\epsfbox{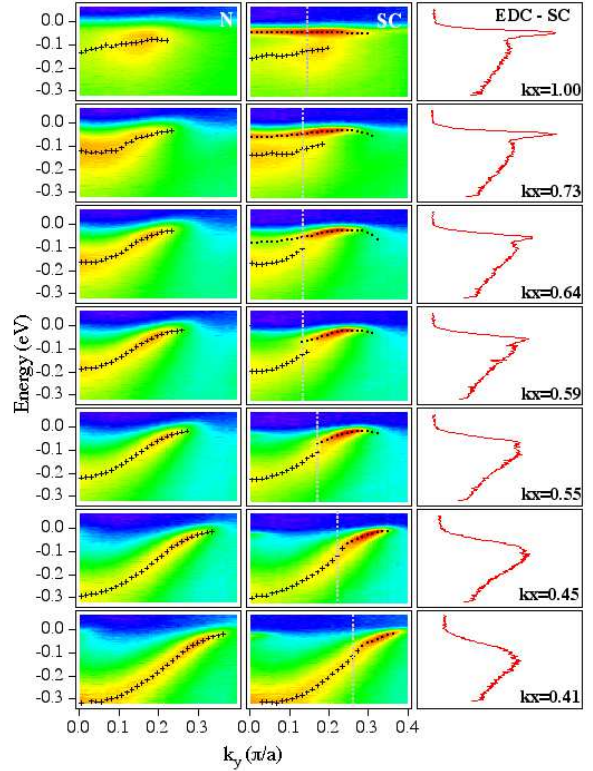}
\vspace{0.2cm}
\caption{
Left panels: Log of normal state ($h\nu$=22eV, T=140K) ARPES intensity
along selected cuts parallel to $MY$.
(shown in zone inset in Fig.~1). EDC peak positions are indicated by
crosses. Middle panels: Log of superconducting state (T=40K) intensity
at the
same cuts as for left panels. Crosses indicate positions of broad high
energy
peaks, dots sharp low energy peaks.  Right panels: EDCs at locations
marked
by the vertical lines in the middle panels.}
\label{fig4}
\end{figure}

\begin{figure*}[!t]
\epsfxsize=6.0in
\epsfbox{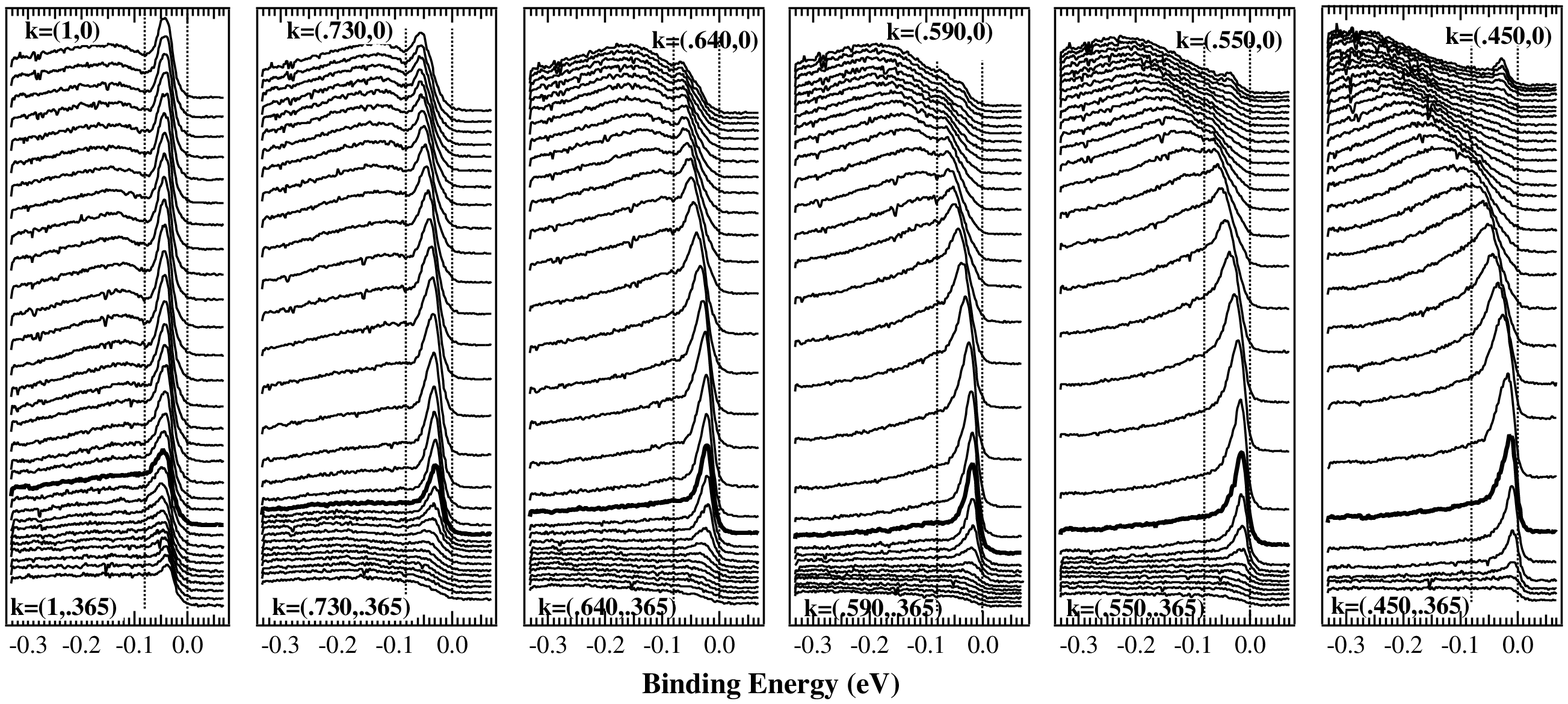}
\vspace{0.2cm}
\caption{
ARPES intensity (T=40K) along selected cuts from Fig.~4. The thick lined
curves
correspond approximately to ${\bf k}_{F}$.  Vertical lines are at 0 and
-80 meV.
}
\label{fig5}
\end{figure*}

The above evolution is suggestive of the self-energy becoming stronger 
as the $(\pi,0)$ point is approached. This can be quantified from the 
observed change in the dispersion. In Fig.~3(e) we plot the ratio of 
the EDC dispersion slope above and below the kink energy at various
points along the Fermi surface obtained from middle panels of Fig.~4.
Near the node, this ratio is around 2, but becomes large near the 
$(\pi,0)$ point because of the nearly dispersionless quasiparticle 
peak\cite{MODE1}.

The lineshape also indicates that the self-energy is larger near 
$(\pi,0)$, as is evident in Fig.~5. Along the diagonal, there is a 
gentle reduction in $\Sigma^{\prime\prime}$ at low energies, as shown 
in Fig.~3 (a) and (b), with an onset at the dispersion kink energy scale.
In contrast, near the $(\pi,0)$ point there must be a very rapid change
in $\Sigma^{\prime\prime}$ in order to produce a spectral dip, as quantified
in Refs.~\cite{MODE1,SELF}. Despite these differences, it is important 
to note that these changes take place throughout the zone at the
same characteristic energy scale (vertical line in Fig.~5).

As discussed in Ref.~\cite{MODE1} the near-$(\pi,0)$ ARPES spectra can be 
naturally explained in terms of the interaction of the electron with a 
collective mode of electronic origin which only exists below $T_c$. It 
was further speculated that this mode was the neutron resonance 
\cite{KEIMER}, an interpretation which received further support from 
Ref.~\cite{MODE2} where the doping dependence of ARPES spectra were 
examined. Here we have shown that dispersion and lineshape anomalies 
have a continuous evolution throughout the zone and are characterized 
by a single energy scale. This leads us to suggest that the same 
electron-mode interaction determines the superconducting lineshape 
and dispersion at all points in the zone, including the nodal direction 
\cite{MIKE}. In essence, there is a suppression of the low
energy scattering rate below the finite energy of the mode.
Of course, since the neutron mode is characterized by a
$(\pi,\pi)$ wavevector, one would expect its effect on the
lineshape to be much stronger at points in the zone
which are spanned by $(\pi,\pi)$ \cite{SS}, as observed here.

In summary, we have shown by a simple, self-consistent analysis based
on general properties of the spectral function and self-energy, that
Bi$_2$Sr$_2$CaCu$_2$O$_{8+\delta}$ shows a dispersion renormalization
along the zone diagonal which is directly related to a drop in the low
energy scattering rate below $T_c$. The anomalies in the dispersion and
lineshape evolve smoothly as one moves from the zone diagonal to the
zone corner, but always show the same characteristic energy scale. We
suggest that this suppression of the scattering rate below $T_c$ at all
points in the Brillouin zone is due to the presence of a gap and a
finite energy collective mode, which we identify with the magnetic
resonance observed by neutron scattering.

MR would like to thank P.D. Johnson for discussions. This work was 
supported by the NSF DMR 9974401, the U.S.~DOE, Basic Energy Sciences, 
under contract W-31-109-ENG-38, the CREST of JST, and the Ministry 
of Education, Science, and Culture of Japan. The Synchrotron Radiation 
Center is supported by NSF DMR 9212658. JM is supported by the Swiss 
National Science Foundation, and MR in part by the Indian DST through 
the Swarnajayanti scheme.

{\it Note added:}  After completion of this work, we became aware
of related work by Bogdanov {\it et al.}, cond-mat/0004349.


\begin{references}
\bibitem{RESOLUTION}
The experimental signal
is a convolution of this
with the energy resolution and a sum over the momentum
window, plus an additive (extrinsic) background.
\bibitem{SIGMA}
Note that: (1) a linear dependence of $\Sigma^{\prime}$ on
$k-k_F$ can be absorbed into the definition of $v_F^0$;
(2) $\Sigma$ has an implicit dependence on $\theta$.
\bibitem{COMPLICATIONS}
Such an analysis would need to include
(1) the superconducting
gap which affects $\Sigma$ through a pairing contribution;
(2) the quadratic dispersion about the $\Gamma M$ symmetry line,
which is close to $E_F$ near $M=(\pi,0)$.
\end{references}
\end{document}